\documentclass[a4paper,11pt]{article}
\pdfoutput=1
\usepackage{jcappub}
\usepackage{slashed}
\usepackage{bm} 
\usepackage{graphicx,epsf,epstopdf}
\usepackage{latexsym,amssymb,amsmath,float,url,mathrsfs}
\usepackage{soul}
\usepackage{natbib}
\usepackage{slashed}
\usepackage{subfigure}

\newcommand{\beq}[1]{\begin{equation}\label{#1}}
\newcommand{\eeq}{\end{equation}}
\newcommand{\bea}{\begin{eqnarray}} 
\newcommand{\eea}{\end{eqnarray}}
\newcommand{\ba}{\begin{array}}
\newcommand{\ea}{\end{array}}

\newcommand{\dmsq}{\delta m^2_\eta}
\newcommand{\vev}{v_{\rm EW}}

\input{epsf}

\begin{document}

\title{\boldmath 
Mono-$W$ Dark Matter Signals at the LHC: Simplified Model Analysis}

\author{Nicole F.\ Bell,}
\author{Yi Cai and}
\author{Rebecca K.\ Leane}
\affiliation{ARC Centre of Excellence for Particle Physics at the Terascale \\
School of Physics, The University of Melbourne, Victoria 3010, Australia}

\emailAdd{n.bell@unimelb.edu.au}
\emailAdd{yi.cai@unimelb.edu.au}
\emailAdd{rleane@physics.unimelb.edu.au}

\date{\today}

\abstract{We study mono-$W$ signals of dark matter (DM) production at the
  LHC, in the context of gauge invariant renormalizable models. 
We analyze two simplified models, one involving an $s$-channel $Z'$
mediator and the other a $t$-channel colored scalar mediator, and
consider examples in which the DM-quark couplings are either isospin
conserving or isospin violating after electroweak symmetry breaking.
While previous work on mono-$W$ signals have focused on isospin
violating EFTs, obtaining very strong limits, we find that isospin
violating effects are small once such physics is embedded into a gauge
invariant simplified model. We thus find that the 8 TeV mono-$W$
results are much less constraining than those arising from mono-jet searches.
Considering both the leptonic (mono-lepton) and hadronic (mono fat
jet) decays of the $W$, we determine the 14 TeV LHC reach of the
mono-$W$ searches with $3000$ $fb^{-1}$ of data.
While a mono-$W$ signal would provide an important complement to a
mono-jet discovery channel, existing constraints on these models imply
it will be a challenging signal to observe at the 14 TeV LHC.}

\maketitle


\section{Introduction}
\label{sec:intro}

Since dark matter (DM) was first suggested over 80 years ago,
compelling evidence has accumulated for its existence across
cosmological scales.  However, the details of the fundamental particle
properties of dark matter remain elusive.  There exist a plethora of
models which provide possible DM candidates, among which a
particularly attractive and well-motivated class is weakly interacting
massive particles (WIMPs) \cite{Bergstrom:2000pn,Bertone:2004pz}.

The exact details of WIMP interactions with Standard Model (SM)
particles are unknown, and it is thus convenient to describe these
interactions in a model-independent manner. This is often done within
an effective field theory (EFT) framework, in which the high energy
renormalizable interactions are approximated at low energy by a set
of non-renormalizable
operators~\cite{Goodman:2010yf,Goodman:2010ku,Duch:2014xda}.  This low
energy description is obtained from the full high-energy theory by
integrating out heavy degrees of freedom.
For fermionic dark matter, $\chi$, interacting with SM fermions, $f$,
the EFT operators take the form:
\begin{equation}
\frac{1}{\Lambda^2}\left( \overline{\chi} \Gamma_\chi \chi \right) \left( \overline{f} \Gamma_f f\right),
\end{equation}
where the remnants of the high energy theory are encapsulated by the
parameter $\Lambda$, which contains the mass $M$ of the mediator and
its couplings $g_i$ in the form $\Lambda=M/\sqrt{g_1g_2}$, and by
$\Gamma_{\chi,f}$, which are the Lorentz structures of the
interaction.

These EFT operators have found wide application in the LHC mono-$X$
searches for DM production
\cite{Aad:2015zva,Aad:2014tda,ATLAS:2014wra,Aad:2014vka,Aad:2013oja,Aad:2014wza,Aad:2014vea,
  Khachatryan:2014rra,Khachatryan:2014rwa,Khachatryan:2014tva,Khachatryan:2014uma,Khachatryan:2015nua,
  Carpenter:2012rg, Carpenter:2013xra, Petrov:2013nia, Bell:2012rg,
  Bai:2012xg, Birkedal:2004xn, Gershtein:2008bf, Goodman:2010ku,
  Crivellin:2015wva, Petriello:2008pu, Berlin:2014cfa, Lin:2013sca,
  Fox:2011pm, Bai:2015nfa, Autran:2015mfa, Gupta:2015lfa}.  These are generic search channels in which a visible
SM final state recoils against the missing momentum carried off by a
pair of DM particles.  Typically the mono-jet channel provides the
most stringent constraints, while mono-$W$, $Z$, $\gamma$ or Higgs
signals would provide indispensable complementary information to
identify DM.

The EFT approximation is valid when the momentum transfer in a given
process of interest is much smaller than the mass of the mediating
particle. For momentum transfer larger than or comparable to
$\Lambda$, the EFT description will break down.  This situation is
likely to arise at the LHC, where the momentum of the partons in the
colliding protons, and thus the momentum transfer of the scattering
processes, will be of TeV scale and comparable to $\Lambda$ in many
WIMP scenarios.  The precise values of the parameters for which this
break down occurs have been the subject of several recent
papers~\cite{Busoni:2013lha,Busoni:2014sya,Busoni:2014haa,Buchmueller:2013dya}.  An
alternative framework which avoids these issues is ``simplified
models''~\cite{Abdallah:2014hon,Buckley:2014fba,Alves:2011wf,Alwall:2008ag, Abdallah:2015ter, Abercrombie:2015wmb}. In
this framework a mediator is explicitly included and interaction types
which are generic yet phenomenologically distinct are considered.

However, the validity of the EFT description is not governed only by
the size of $\Lambda$.  The standard list of EFT
operators~\cite{Goodman:2010yf,Goodman:2010ku} include several which
do not respect the weak gauge symmetries of the SM\footnote{Indeed,
  some simplified models also have this shortcoming.}.  Such operators
break down at the energies comparable to the electroweak scale,
$\vev\approx 246$~GeV, rather than the energy scale of new physics,
$\Lambda$, and are certainly invalid at LHC energies.  In fact, such
operators should be suppressed by powers of $(\vev/\Lambda)^n$, and
are thus of higher order in $1/\Lambda$ than they naively appear.  One
should proceed with caution in interpreting LHC limits on such
operators.

In a recent paper~\cite{Bell:2015sza} we demonstrated that operators which
violate weak gauge symmetries can feature spurious cross section
enhancements at LHC energies.  This was particularly pertinent for
previous mono-$W$ searches for dark matter at the LHC \cite{Khachatryan:2014tva, ATLAS:2014wra, Aad:2013oja}, which have
largely focused on $SU(2)$ violating EFTs such as \cite{Bai:2012xg}
\begin{equation}
\frac{1}{\Lambda^2}\left( \overline{\chi} \gamma^\mu \chi \right) \left( \overline{u} \gamma_\mu u + \xi \overline{d} \gamma_\mu d\right),
\label{eq:xi}
\end{equation}
with $\xi \neq 1$.  The large mono-$W$ cross sections for such an EFT
are in fact a manifestation of the violation of weak gauge invariance
in the form of unphysical longitudinal $W$ contributions.  Previous
work has used these unphysical enhancements of the mono-$W$ cross
section to place very strong limits on dark matter EFTs.  However,
when gauge invariance is enforced we shall see that the limits arising
from the mono-$W$ process will in general be weaker than those
arising from the mono-jet.  Nonetheless, the mono-$W$ process remains an
important complementary channel to explore the properties of dark matter.

It is the purpose of the present paper to study mono-$W$ signals in
renormalizable models in which gauge invariance is enforced from the
outset.  We choose two example simplified models, one involving
$t$-channel exchange of a new colored scalar, and the other $s$-channel
exchange of a new $Z'$ vector boson.  We outline these two models in
Section~\ref{sec:simplified}.  In Section~\ref{sec:lhc} we explore the
LHC phenomenology of these models, to determine the current
constraints and the 14 TeV LHC reach for the mono-$W$ signal.  In
Section~\ref{sec:wl} we explore the possibility of obtaining $SU(2)$
violating operators, like that of Eq. (\ref{eq:xi}), from a gauge
invariant model after electroweak symmetry breaking.  While such
operators would allow for the production of longitudinal $W_L$ bosons,
potentially enhancing mono-$W$ cross sections, we explain why these
effects are constrained to be small.

\section{Simplified Models for the Mono-$W$}
\label{sec:simplified}

\subsection{$t$-channel Colored Scalar Mediator}

We first examine a scenario in which DM-quark interactions are
mediated by the exchange of a $t$-channel scalar.  The interaction Lagrangian
is given by
 \begin{eqnarray}
\mathcal{L}_\text{int}&=&f \overline{Q}_L \eta \chi_R+h.c.
\nonumber\\ 
&=&f_{ud}\left(\eta_u\overline{u}_L+\eta_d\overline{d}_L\right)\chi_R+h.c.,\label{eq:tchannel}
\end{eqnarray}
where $Q_L = (u_L,d_L)^T$ is the quark doublet,
$\eta=(\eta_u,\eta_d)^T \sim (3,2,1/6)$ is a scalar field that
transforms under the SM gauge group like $Q_L$, and $f$ is the
coupling strength of the interactions\footnote{One can write down a
  similar model involving a coupling to right handed (RH) quark fields.  While most
  of the phenomenology would be very similar, such a model would not
  permit a mono-$W$ signal. Isospin violating models with RH quark fields were considered in \cite{Feng:2011vu, Feng:2013vod}.}.  The DM, $\chi$, transforms as
a singlet under the SM gauge symmetries. An analogue of this scenario
is realized in supersymmetric (SUSY) models, where we identify $\eta$
with a squark doublet and $\chi$ the neutralino.  Simplified models
with such $t$-channel interactions have been examined recently in
Refs.~\cite{Difranzo:2013, Bai:2013iqa, Wang:2014, Chang:2014, Papucci:2014iwa, Garny:2015wea, Garny:2014waa}, with
the collider analyses focusing on the mono-jet process.

\begin{figure*}[ht]
\centering
\subfigure[]{\includegraphics[width=0.3\columnwidth]{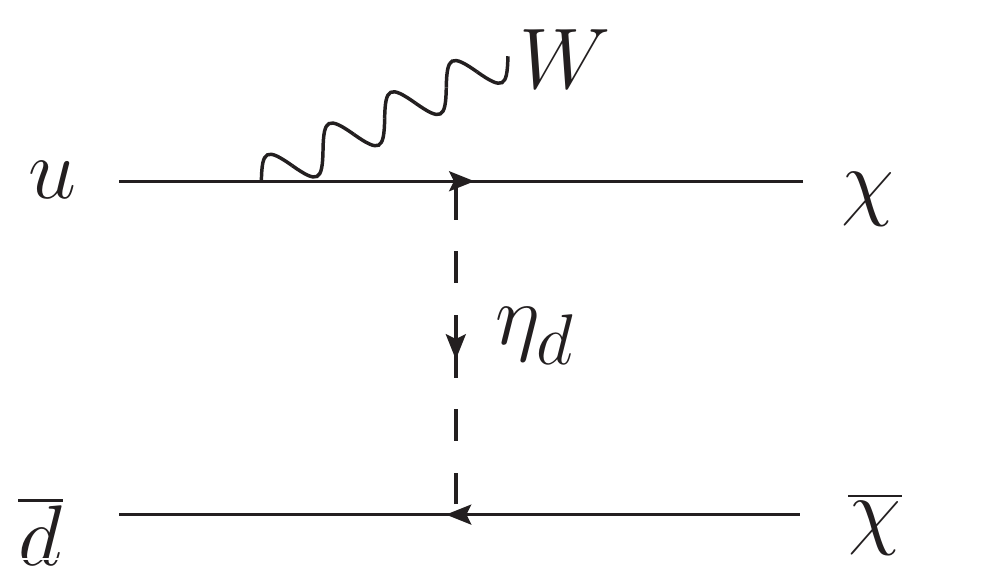}}
\subfigure[]{\includegraphics[width=0.3\columnwidth]{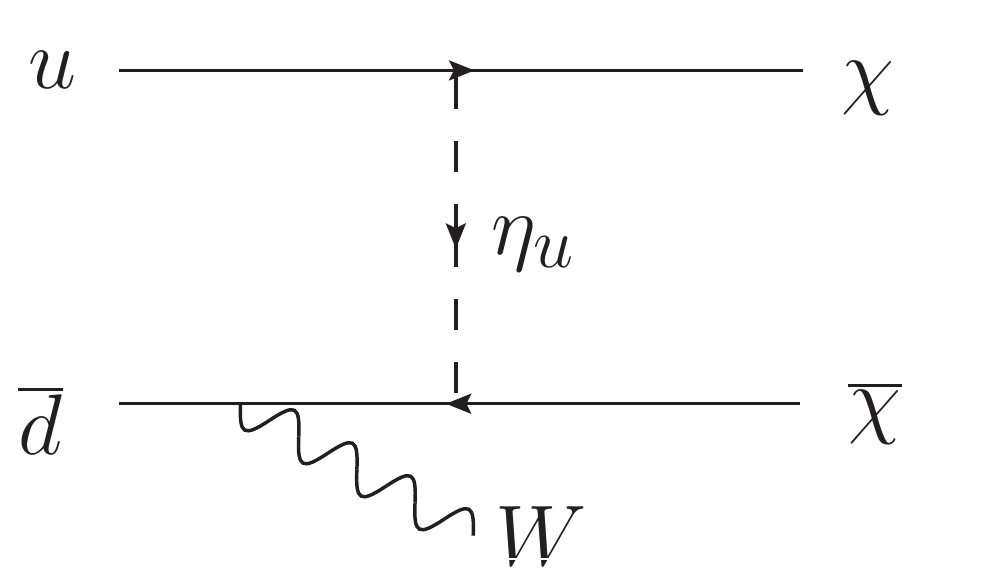}}
\subfigure[]{\includegraphics[width=0.3\columnwidth]{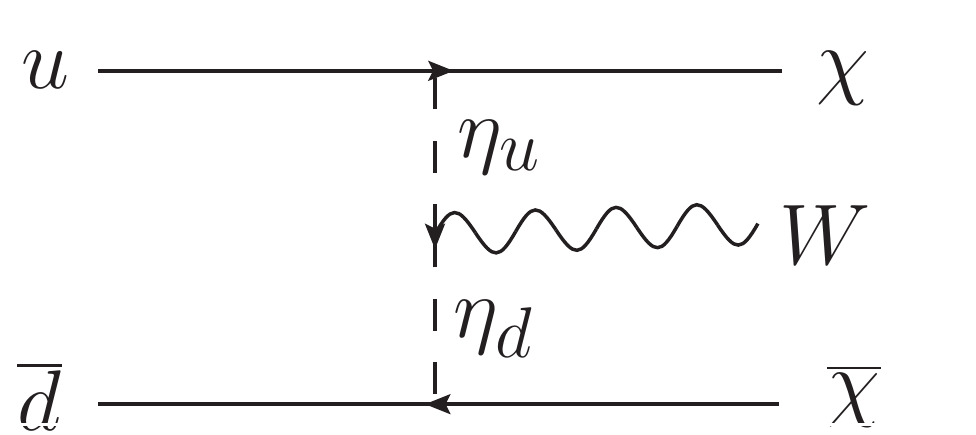}}
\caption{Contributions to the mono-$W$ process $ u(p_1)\overline{d}(p_2)  \rightarrow  \chi(k_1)\overline{\chi}(k_2)  W^+(q)$, in a $t$-channel colored scalar model.}
\label{fig:diags}
\end{figure*}

In this model, the mono-$W$ process proceeds via the gauge invariant
set of diagrams in Fig. (\ref{fig:diags})
~\cite{Bell:2011if,Ciafaloni:2011sa,Bell:2012rg,Bell:2015sza}. Diagrams
(\ref{fig:diags}a) and (\ref{fig:diags}b) dominate in the EFT limit when $\sqrt{s}\ll m_\eta$, while
diagram (\ref{fig:diags}c) becomes important for smaller $m_\eta$.  We shall initially assume $m_{\eta_u} = m_{\eta_d} = m_\eta$.  
Deviation from this equality will be discussed in Section~\ref{sec:wl}.

\subsection{$s$-channel $Z'$ Mediator}

We also consider another generic simplified model in which the DM-quark
interactions are mediated by a neutral spin-1 $Z'$ boson.  The
relevant interaction terms are
 \begin{equation}
\mathcal{L}_\text{int} \supset g_{\chi}\overline{\chi}\gamma^\mu\gamma^5\chi Z'_{\mu} + g_q\overline{q}\gamma^\mu\gamma^5 q Z'_{\mu},
\label{eq:schannel}
\end{equation}
where $g_{\chi}$ is the coupling strength of the $Z'$ to dark matter
$\chi$, and $g_{q}$ is the coupling to SM quarks.  Simplified models
with such $s$-channel interactions have been examined recently in
Refs.~\cite{Buchmueller:2014yoa,Heisig:2015ira,Hooper:2014fda,Blennow:2015gta,
  Lebedev:2014bba, Alves:2015pea, Alves:2013tqa, Alves:2015mua,
  An:2012va, An:2012ue, Frandsen:2012rk, Arcadi:2013qia,
  Shoemaker:2011vi, Frandsen:2011cg, Gondolo:2011eq,
  Fairbairn:2014aqa, Harris:2014hga}.  We assume the $Z'$ has axial
vector type interactions.  Vector interactions would lead to large
spin-independent DM-nucleon elastic scattering cross sections, and as
a result are strongly constrained by DM direct detection experiments,
to the extent that parameters which can correctly account for the DM
relic density are significantly excluded.  We therefore focus on the
more phenomenologically viable axial vector interactions.  We shall
also assume that the $Z'$ couples only to quarks, and not to leptons,
to avoid tight constraints from di-lepton searches.

\begin{figure*}[ht]
\centering
\subfigure[]{\includegraphics[width=0.33\columnwidth]{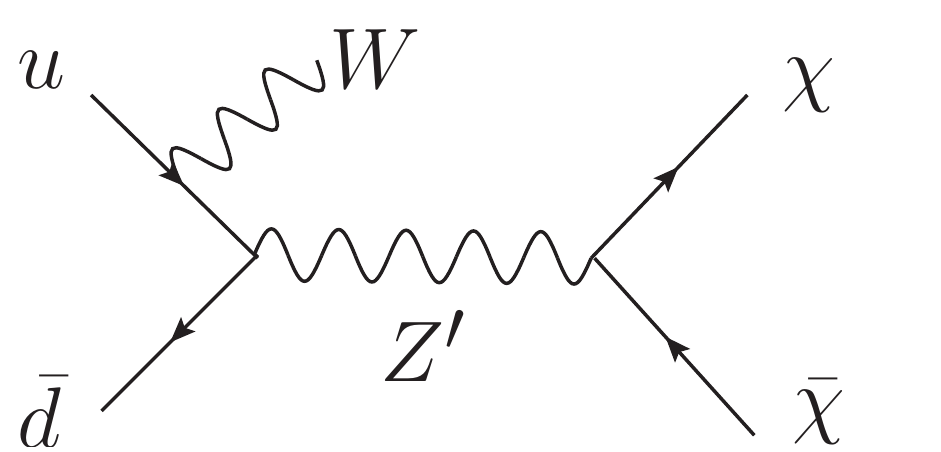}}
\subfigure[]{\includegraphics[width=0.33\columnwidth]{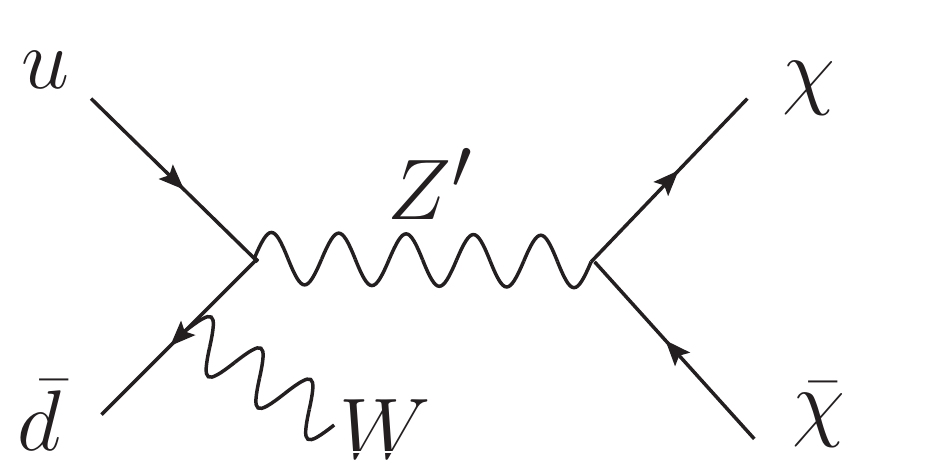}}
\caption{Contributions to the mono-$W$ process $ u(p_1)\overline{d}(p_2)  \rightarrow  \chi(k_1)\overline{\chi}(k_2)  W^+(q)$, in an $s$-channel $Z'$ model.}
\label{fig:diags1}
\end{figure*}

The pertinent processes for mono-$W$ search are shown in
Fig.~(\ref{fig:diags1}).  In contrast to the $t$-channel model above, no
radiation from the mediator occurs.  This would change in the presence
of $Z$-$Z'$ mixing, as will be discussed in Section \ref{sec:wl}.

\section{LHC Constraints and Reach}
\label{sec:lhc}

We now examine the LHC phenomenology of the two models described in
Eqs. (\ref{eq:tchannel}) and (\ref{eq:schannel}).  In the following, we
determine the limits and reach of the searches for DM via the mono-$W$
process, for both the leptonic and hadronic decay channels of the $W$.

\subsection{Mono lepton channel}

We first consider the scenario where the $W$ boson decays to a charged lepton and a neutrino. The neutrino contributes to the missing energy ($\slashed{E}_T$) along with dark matter, such that the signal is a mono-lepton. In this channel
the key kinematic variable is the transverse mass of the lepton-$\slashed{E}_T$ system,
\begin{equation}
 M_T=\sqrt{2p_T^\ell \slashed{E}_T(1-\cos\Delta\phi_{\ell, \slashed{E}_T})} \; ,
\end{equation}
where $\Delta\phi_{\ell,\slashed{E}_T}$ is the azimuthal opening angle between the charged lepton's transverse momentum
$p_T$ and the direction of  $\slashed{E}_T$.

The domininant background for the mono-lepton search is $W\rightarrow \ell\nu$,  and $W\rightarrow\tau\nu_\tau\rightarrow \nu_\tau \nu_\tau \ell \nu_\ell$ where $\ell = e, \mu$. This is because the $M_T$ distribution of these channels has a large tail in the signal region.
We use the electron channel to set limits, since it is the stronger one of two lepton channels and also comparable to the combined limits of both channels. 
Following Ref.  \cite{Khachatryan:2014tva}, the following selection cuts are made on all backgrounds and signal for the 8 TeV limits:

\begin{itemize}
  \item $E_T$ of the leading electron $>$ 100 GeV
  \item $E_T$ of the next-to-leading electron $<$ 35 GeV
  \item At least one electron 
  \item $M_T$ for the electron, $M_T^e > 220$ GeV
  \item Pseudorapidity for the electron must be in the range $-2.1 < \eta(\ell_e) < 2.1$
  \item Jet $P_T < 45$ GeV
  \item The electron $P_T$ and $\slashed{E}_T$ ratio must be in the range $0.4 < P_T/\slashed{E}_T < 1.5$
  \item $\Delta\phi_{e,\slashed{E}_T} > 2.5$.
\end{itemize}

\begin{figure}[H]
  \centering
  \includegraphics[width=10cm]{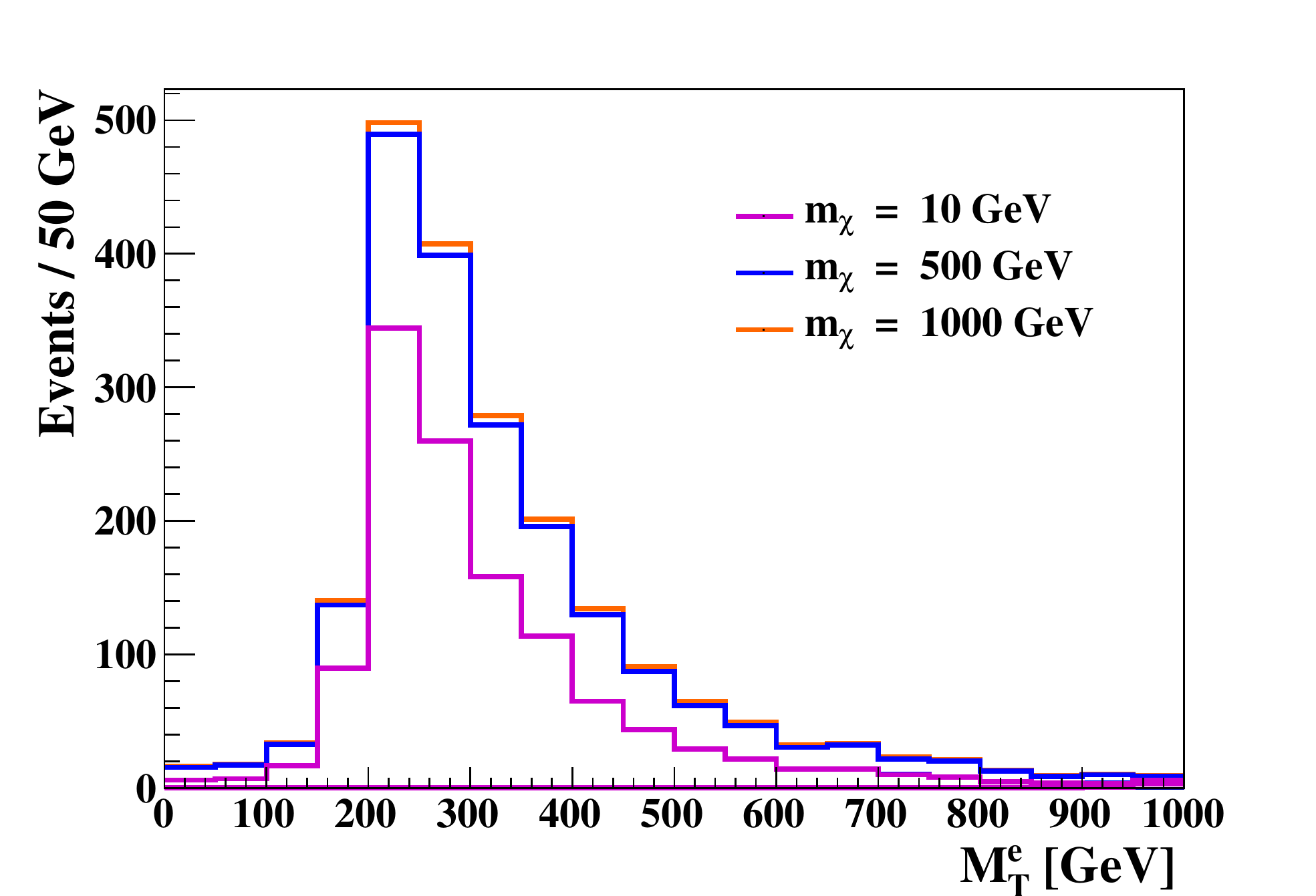}
  \caption{$M_T$ distribution for $m_{\eta_u}=m_{\eta_d}=200$ GeV, $g=1$, $m_\chi = 10, 500, 1000$ GeV in the $t$-channel model, at 14 TeV and $\mathcal{L}_{int}=3000$ $fb^{-1}$. It can be seen that the distribution is independent of DM mass.}
\label{Fig:MTmin}
\end{figure}

After cuts, the events are scaled by the relevant efficiences. To investigate the phenomenology, both models are implemented in {\sc FeynRules} \cite{Degrande:2011ua}. For the mono-lepton search, events are generated in {\sc MadGraph/MadEvent} \cite{Alwall:2011uj, Maltoni:2002qb}, hadronized in {\sc Pythia} \cite{Sjostrand:2006za}, interfaced with {\sc Fastjet} \cite{Cacciari:2011ma} for jet-finding and
{\sc Delphes} \cite{deFavereau:2013fsa} for detector effects. We then implement our cuts in {\sc Root} \cite{Brun:1997pa}, and set the model significance $\sigma$ at 95 $\%$ confidence level (C.L.), which is set by the number of signal events $S$ and background events $B$ as
\begin{equation}
 \sigma=\frac{S}{\sqrt{S+B+(\delta B ^*B)^2}} \; ,
\label{eq:sigma}
\end{equation}
where $\delta B$ is the systematic uncertainty, which we take to be
$5\%$ for our analysis. To ensure a thorough sampling of events and
sufficient statistics at high $M_T$, we generate event samples at two
different regions for both signal and background, $100 < p_T^\ell <
500$ GeV, and $p_T^\ell > 500$ GeV. The samples from these two regions
are then combined to produce the background and signal events.  We
find that we reproduce the model independent limit on the cross
section for a mono-lepton signal as found in
Ref.~\cite{Khachatryan:2014tva}, at 8 TeV.  We then perform the
analysis at 14 TeV and 3000 $fb^{-1}$ integrated luminosity. To
produce our 95$\%$ C.L. reach, we optimize our 14 TeV selection
criteria by increasing the $M_T$ cut to $M_T^e > 1000$ GeV.  In
Fig. (\ref{Fig:MTmin}) we show the $M_T$ distribution for the $t$-channel
model for various choices of the DM mass.  (Similar results are found
for the $s$-channel model.)  As the shape of the $M_T$ distribution is
approximately independent of the DM mass, we adopt $M_T > 1000$ GeV as
an optimal selection cut across all parameter choices.

\subsection{Mono fat jet channel}

We also consider the limits and reach from the hadronic $W$ decays.
Such decays have been searched for by ATLAS \cite{Aad:2013oja}, where
the signal is a hadronically decaying $W$ or $Z$ boson plus missing
energy.  As our simplified models allow both mono-$W$ and mono-$Z$
processes, both must be included in our generated signal.  We refer to
this channel as the ``mono fat jet'' channel as the hadronic decay
products $jj$ of the $W/Z$ can be strongly boosted such that they
appear together as one wide jet, making the signal this ``fat jet''
along with $\slashed{E}_T$ from DM.

The relevant backgrounds for this search are $Z\rightarrow\nu\bar{\nu}$, $W\rightarrow\ell^{\pm}\nu$, $Z\rightarrow\ell\ell$, $WW$ $WZ$ $ZZ$, $t\bar{t}$ and top production. 
We generate backgrounds in {\sc Herwig++} \cite{Bahr:2008pv}, where events are also hadronized. Using both our models implemented in {\sc FeynRules} \cite{Degrande:2011ua}, 
signal events are generated in {\sc MadGraph/MadEvent} \cite{Alwall:2011uj, Maltoni:2002qb} and are hadronized in {\sc Pythia} \cite{Sjostrand:2006za}. Both signal
and background events are then passed to external {\sc Fastjet} \cite{Cacciari:2011ma}, where we implement jet finding algorithms and cuts, followed
by {\sc Delphes} \cite{deFavereau:2013fsa} for detector effects and efficiencies. Specifically, in order to discriminate between background jets and those produced by the $W/Z$,
a mass-drop filtering procedure is used. Here, large radius jet candidates which mostly come from the decay of the $W/Z$ are first reconstructed via the Cambridge-Aachen algorithm~\cite{CMS:2009lxa} with a radius parameter of 1.2. Then, the internal structure of this
large radius jet is examined, and the subjets, called ``narrow jets'', are reconstructed using the anti-kt jet clustering algorithm \cite{Cacciari:2008gp} with a radius parameter of 0.4. The mass-drop is performed on the two leading subjets, where the subjet with the largest
 $p_T$, $p_{T1}$ differs from the momentum of the next to leading subjet $p_{T2}$ by
 \begin{equation}
 \sqrt{y}=min(p_{T1},p_{T2})\frac{\Delta R}{m_{jet}}, 
 \end{equation}
 where $\Delta R$ is the separation of the two leading subjets and $m_{jet}$ is the mass of the large radius jet. For 8 TeV, following the analysis of \cite{Aad:2013oja}, we also require:

\begin{itemize}
  \item $\slashed{E}_T$ $>$ 350 GeV
  \item At least one large radius jet with $P_T > 250$ GeV
  \item $\sqrt{y}>0.4$
  \item $50 < m_{jet} < 120$ GeV
  \item $-1.2 < \eta < 1.2$
  \item No more than one narrow jet with $P_T > 40$ GeV and $-4.5 < \eta < 4.5$ which
  is separated from the leading large radius jet as $\Delta R > 0.9$
  \item $\Delta\phi(jet,\slashed{E}_T) < 0.4$ for narrow jets.
\end{itemize}

As the $Z\rightarrow\nu\nu$ background process in this channel has low statistics after cuts, to ensure a thorough probe of phase space we generate and average 6 sets of 50,000 events at 14 TeV for this background. 
For the other background processes, we generate 50,000 events per process. We set the model significance at 95 $\%$ C.L. , as per Eq. (\ref{eq:sigma}).
For the 14 TeV reach, we optimize the search by adjusting three of the 8 TeV selection criteria; we now require at least one large radius jet
with $P_T > 400$ GeV, require $\slashed{E}_T$ $>$ 500 GeV and $70 < m_{jet} < 90$ GeV.

\subsection{Results}

\begin{figure}[H]
  \centering
  \includegraphics[width=10cm]{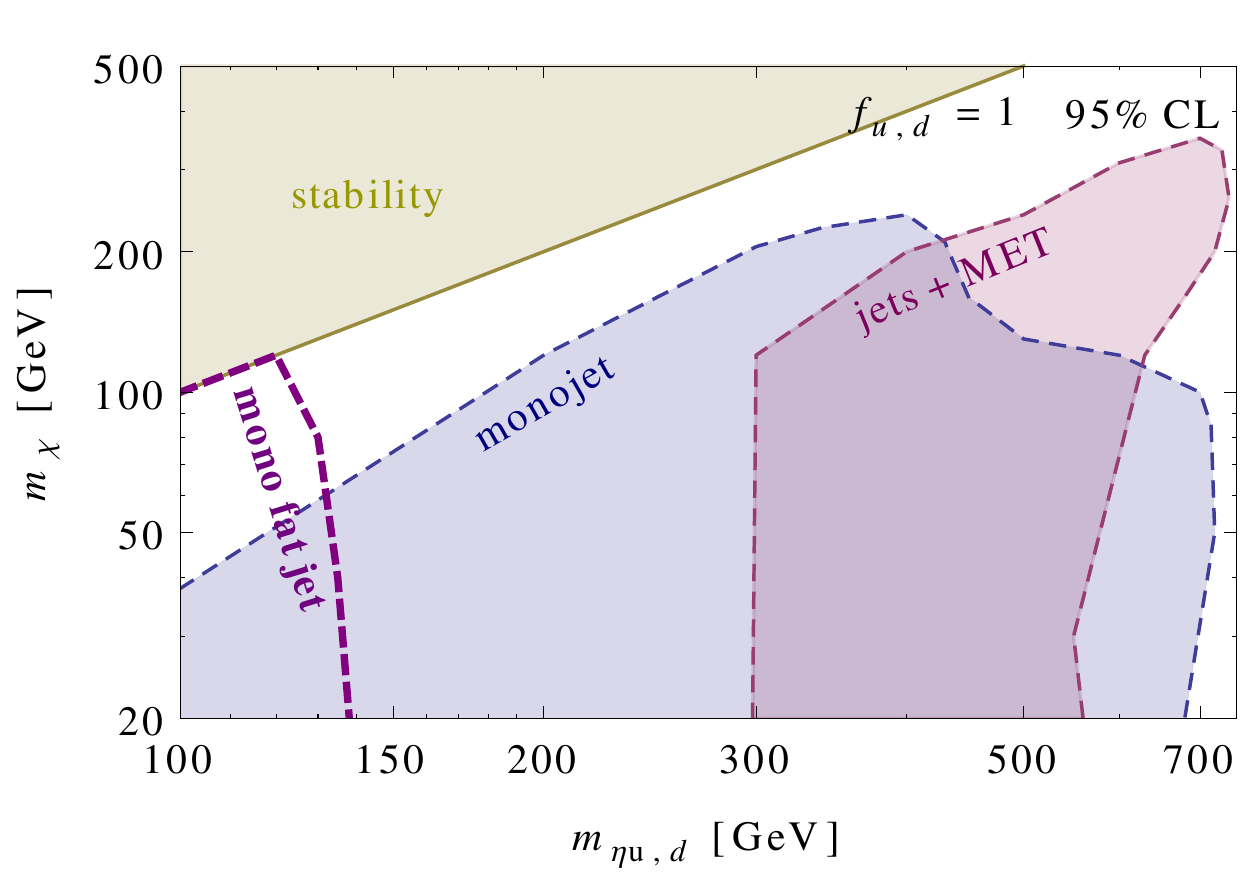}
  \caption{Parameter space for the $t$-channel colored scalar model, for $f_{u,d}=1$. Exclusions are shown as shaded regions for the mono and multi jet at 8 TeV, and the reach is shown for the mono fat jet at 14 TeV 3000 $fb^{-1}$.}
  \label{fig:tchannelplots}
\end{figure}

For the $t$-channel model, the current limits are compared with the 14
TeV mono-$W$ reach in Fig. (\ref{fig:tchannelplots}) for $f_{u,d}=1$.
We also include current constraints on the parameter space from
mono-jet and multi-jet searches, which are adopted from
Ref.~\cite{Bai:2013iqa}.  The region labelled ``stability'' is
forbidden as it corresponds to parameters where $m_\chi>m_\eta$ and
thus the DM would be unstable to decay.
For the mono-lepton search, we find that both the current 8 TeV
exclusion and 14 TeV reach are not competitive with existing
constraints from mono-jet searches.  Owing to small signal size and
large backgrounds, it is too weakly constraining to be featured on our
$t$-channel summary plot.  For the mono fat jet search, we find that the
8 TeV exclusions are also not competitive with existing constraints
from mono-jet searches.  We show the 14 TeV reach in the mono fat jet
channel with 3000 $fb^{-1}$ of data, which is able to probe a region
of parameter space unconstrained by existing mono-jet results.

\begin{figure*}[h]
\centering
\vspace{-4mm}
\subfigure[]{\includegraphics[width=9.0cm]{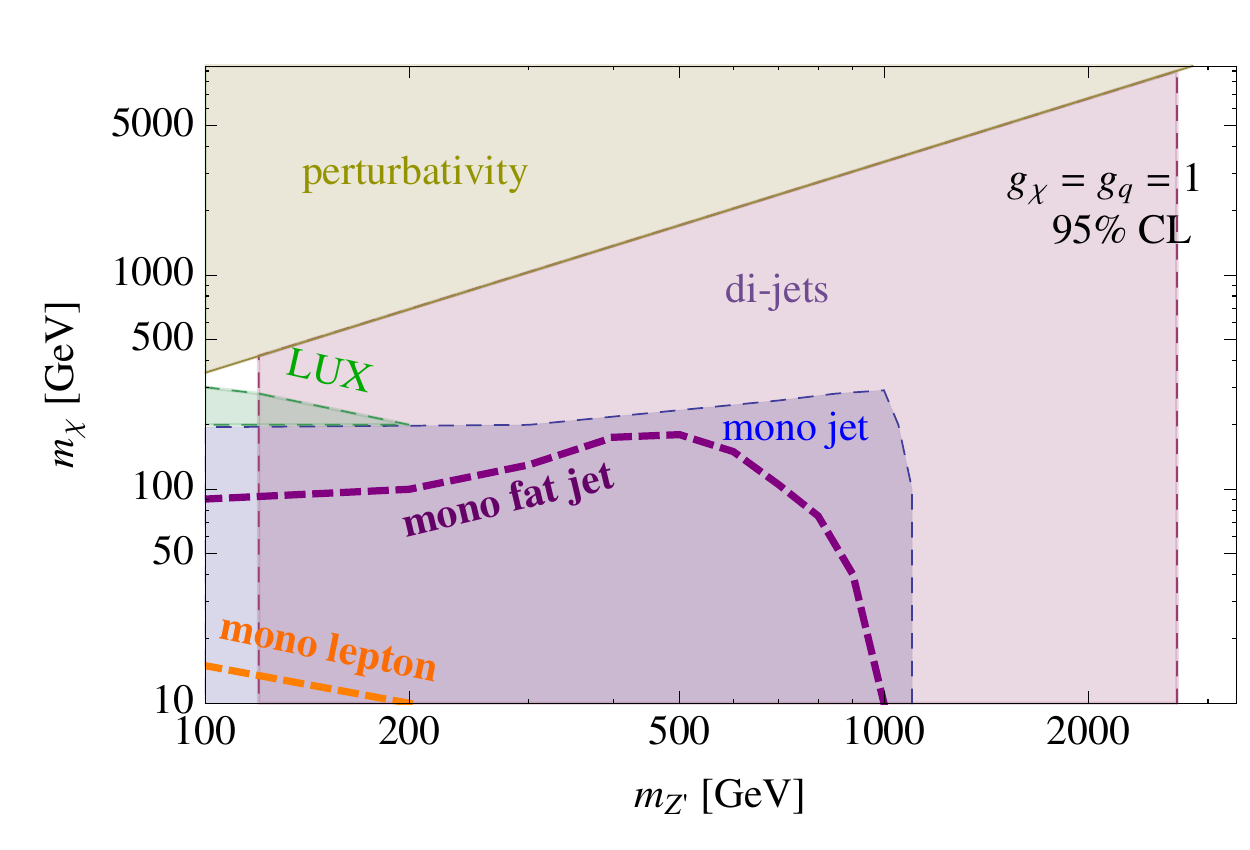}}\vspace{-3mm}
\subfigure[]{\includegraphics[width=9.0cm]{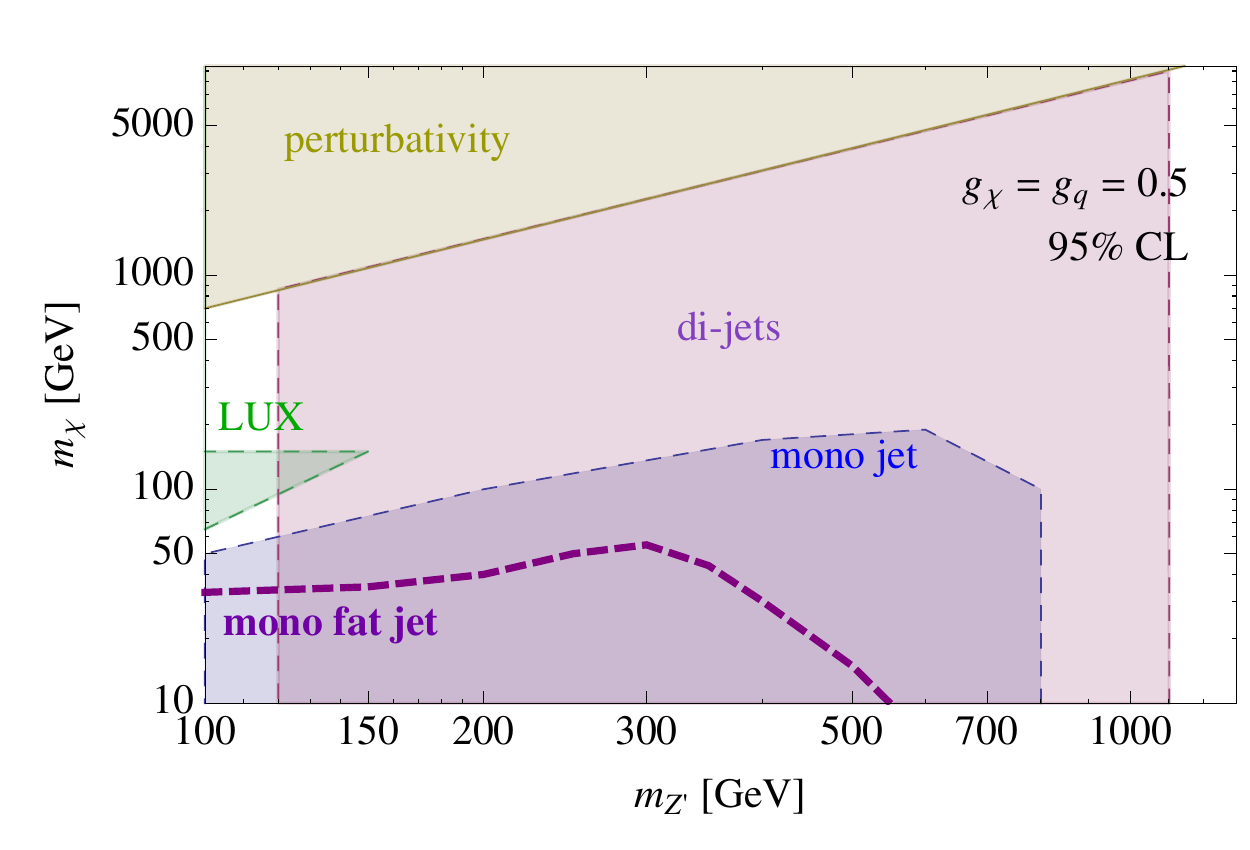}}\vspace{-3mm}
\subfigure[]{\includegraphics[width=9.0cm]{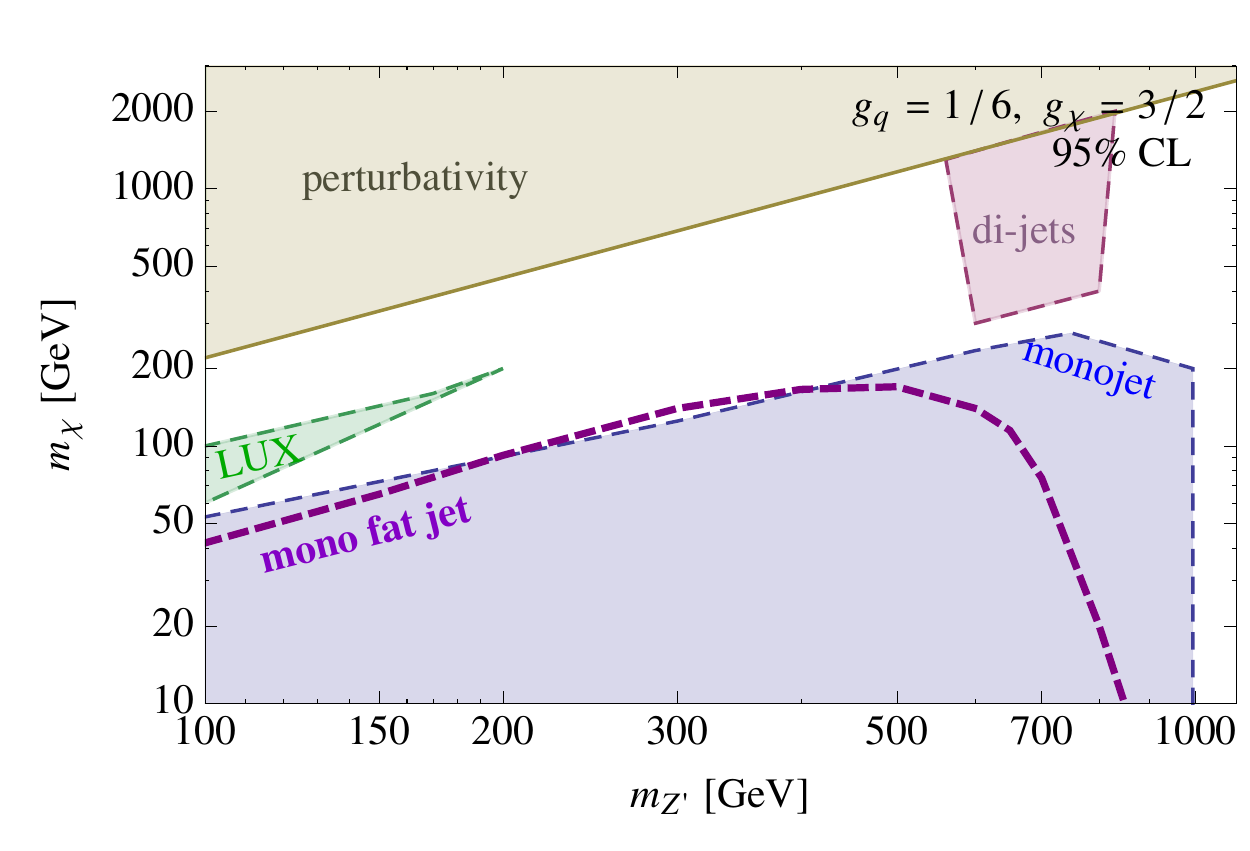}}
\caption{Parameter space for the $s$-channel $Z'$ model, for choices of (a) $g_{q}=g_{\chi}=1$ and (b) $g_{q}=g_{\chi}=0.5$ and (c)  $g_{q}=1/6$ and $g_{\chi}=3/2$. Exclusions are shown as shaded regions for LUX and for mono-jet and di-jets at 8 TeV, and the reaches are shown for the mono lepton ((a) only) and mono fat jet searches at 14 TeV 3000 $fb^{-1}$.
Note differing axes.}\vspace{-2mm}
\label{fig:schannelplots}
\end{figure*}

For the $s$-channel model, our results are shown in
Fig. (\ref{fig:schannelplots}) for three choices of the $Z'$ couplings
to DM and quarks, as labelled.  The relevant mono-jet, di-jet and LUX
\cite{Akerib:2013tjd} direct detection limits shown are adopted from
Ref.~\cite{Chala:2015ama}.  Note that the LUX limit assumes the actual
(sub-critical) contribution to the DM relic density implied by the
model parameters, rather than assuming a full relic density.
We also include perturbativity limits for the $s$-channel model. As has
been recently shown in \cite{Chala:2015ama, Kahlhoefer:2015bea}, the
$s$-channel model with axial couplings may have perturbativity and
unitarity issues without the inclusion of additional new physics such
a dark Higgs scalar which generates the DM and $Z'$ mass.
Perturbative unitarity implies that the $Z'$ cannot be much lighter
than the DM, and should satisfy
$m_{\chi}\lesssim\frac{\sqrt{4\pi}}{g_\chi}m_{Z'}$.  This is shown on
the $s$-channel plots as the perturbativity region.  While this is not a
concrete exclusion, it is an important issue for this region of
parameter space.

For the mono-lepton search, the current 8 TeV exclusion is too weak to
be shown on the plots, while the 14 TeV reach is shown only for
$g_{q}=g_{\chi}=1$, as it is very weakly constraining for the other
coupling choices. As with the $t$-channel model, the mono fat jet
channel has better sensitivity than the mono-lepton channel, and the 14 TeV
reach is shown for each of the coupling choices.  However, even with
the hadronic decay mode, the mono-$W$ signals will be challenging to
observe, with the parameter space accessible at 14 TeV already
substantially probed by 8 TeV mono-jet searches.

\section{$SU(2)$ Breaking Effects and Enhancements from $W_L$ Production}
\label{sec:wl}

Previous work on the mono-$W$ signal has focused primarily on EFT
operators that violate $SU(2)_L$.  The strong constraints on these
models were shown to arise from unphysical high-energy contributions
from longitudinally polarized $W$ bosons, a manifestation of the lack
of gauge invariance~\cite{Bell:2015sza}.  The strength of the limits
on these $W_L$ dominated processes arose from two effects:
\begin{itemize}
\item 
enhancement of the cross section, due to a leading $s/m_W^2$
dependence for large $s$ (arising from the $W_L$ contribution to the
polarization sum)~\cite{Bell:2015sza} and
\item
a harder $M_T$ distribution~\cite{Khachatryan:2014tva}, which allowed
better separation of signal and background.

\end{itemize}
By contrast, the gauge invariant simplified models that we considered
above, which feature only transverse $W_T$ contributions in the high
energy limit, do not benefit from these effects.  However $SU(2)$
violating effects, such as the unequal coupling of DM to $u$ and $d$
type quarks of Eq. (\ref{eq:xi}), can be generated at higher order by
electroweak symmetry breaking.  This would permit some high energy
$W_L$ contributions to the mono-$W$ process, potentially leading to
stronger constraints.  We analyze the size of such effects in
variations of our simplified models, and show that it is always small.

\subsection{Isospin Violation in the $t$-channel Model}

In the $t$-channel model, the DM interaction with the $u$ and $d$ quarks
can be of unequal strength if the masses of the respective mediators,
$\eta_u$ and $\eta_d$, are non-degenerate.  Inspection of the scalar
potential reveals that this situation can be realised once the SM
Higgs field gains a vev.  The scalar potential is~\cite{Garny:2011cj}
\begin{equation}
V = m_1^2 (\Phi^\dagger \Phi) + \frac{1}{2}\lambda_1 (\Phi^\dagger \Phi)^2 + m_2^2 (\eta^\dagger \eta) + \frac{1}{2}\lambda_2 (\eta^\dagger \eta)^2 + \lambda_3 (\Phi^\dagger \Phi) (\eta^\dagger \eta) +\lambda_4 (\Phi^\dagger \eta) (\eta^\dagger \Phi),
\label{eq:potential}
\end{equation}
where $\Phi$ is the SM Higgs and $\lambda_n$ are coupling constants.
In the case where $m_1^2<0$ and $m_2^2>0$, the SM Higgs doublet
obtains a vev, while $\eta$ does not.  After electroweak symmetry
breaking, a non-zero value of $\lambda_4$ would split the $\eta$ masses as
\begin{eqnarray}
&& m^2_{\eta_d}=m_2^2 + (\lambda_3 +\lambda_4)v_{\textrm EW}^2, \\
&& m^2_{\eta_u}=m_2^2 + \lambda_3 v_{\textrm EW}^2\,,
\end{eqnarray}
so that 
\begin{equation}
\dmsq \equiv m^2_{\eta_d} - m^2_{\eta_u} = \lambda_4\,v_{\textrm EW}^2.
\label{eq:splitting}
\end{equation}
So we appear to have broken the degeneracy of the DM interactions
with $u$ and $d$ type quarks, as in the EFT of Eq. (\ref{eq:xi}).  Does
this indeed allow for $W_L$ production, and how can this be
understood?

It is instructive to appeal to the Goldstone boson equivalence theorem
to understand where $W_L$ production arises.  In the high energy
limit, we may replace $W_L$ with the corresponding Goldstone boson
that (in unitary gauge) provides the gauge boson mass, i.e., we replace
$W_L^+$ with $\phi^+$.  Now consider the 3 diagrams contributing to
the mono-$W$ process shown in Fig. (\ref{fig:diags}).  The $\phi^+$
couples to the quarks with strength given by the quark Yukawa
constants, which vanish in the limit that the quarks are massless.
Under these conditions, there is no $W_L$ contributions from the
diagrams of Fig. (\ref{fig:diags}a) and (\ref{fig:diags}b).

We now turn to the diagram of Fig. (\ref{fig:diags}c) in which the $W$
is radiated from the $\eta$ mediator.  In general, this diagram will
feature both $W_T$ and $W_L$ contributions. From inspection of the
$\lambda_4$ term in Eq. (\ref{eq:potential}), we deduce that $\phi^+$
will couple to $\eta$ according to~\cite{Garny:2011cj}
\begin{equation}
v_{\textrm EW}\lambda_4\eta_d\eta_u^*\phi^+ + h.c.,
\end{equation}
and thus the size of the $\eta_d\eta_u^*W^+_L$ vertex is determined
by $\lambda_4$.  Therefore, switching on $\lambda_4\neq 0$ and hence
$\delta m^2_\eta\neq 0$ opens a $pp\rightarrow\chi\chi W_L$ channel
that does not suffer from suppression by the quark Yukawas.
(By contrast, in the example studied in Section~\ref{sec:lhc} where
where $\lambda_4=0$ and $\delta m^2_\eta=0$, we expect that the high
energy regime will feature only transversely polarized $W$-bosons,
$pp\rightarrow\chi\chi W_T$.)

\subsubsection{Cross Section Enhancement from $W_L$ Contribution}
We have seen that the amplitude for $W_L$ production at high energy is
controlled by $\lambda_4$.  However, $\lambda_4$ also increases the
mass splitting, making $\eta_d$ heavier than $\eta_u$.  Therefore, increasing
$\lambda_4$ will suppress the contribution of
Fig. (\ref{fig:diags}a) due to the heavier $\eta_d$ propagator, while
enhancing the contribution of Fig. (\ref{fig:diags}c) due to $W_L$
production.  The former effect dominates for small values of
$\lambda_4$, while the latter compensates or dominates if $\lambda_4$
is sufficiently large.

\begin{figure}[H]
  \centering
  \includegraphics[width=10cm]{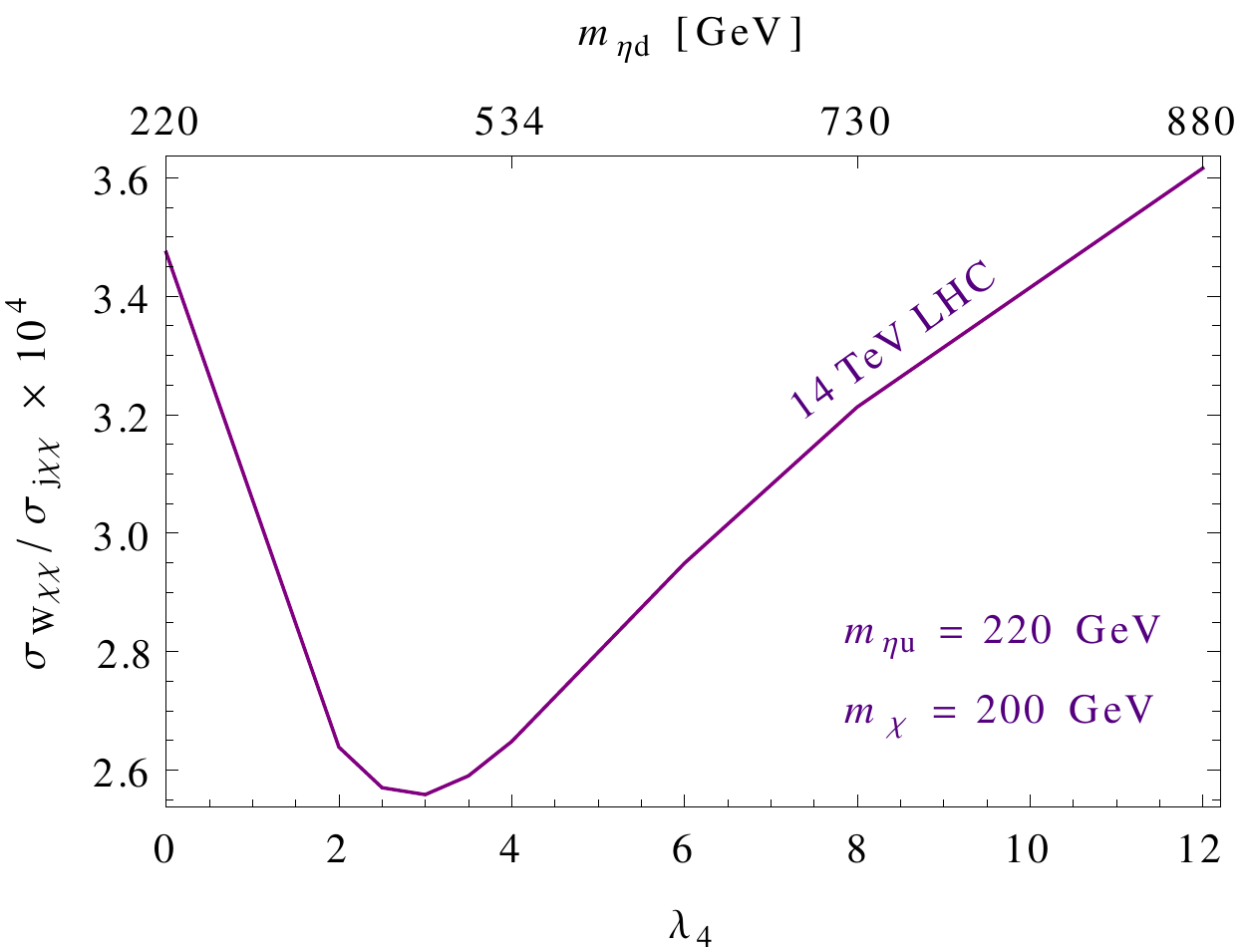}
  \caption{Ratio of the hadron level cross sections for the mono-$W$ process $ pp  \rightarrow  \chi\overline{\chi}  W$, $\sigma_{W\chi{\chi}}$, to the mono-jet process $pp \rightarrow \chi{\chi}j$, $\sigma_{j\chi{\chi}}$  at 14 TeV, in a renormalizable $t$-channel scalar model with isospin violation. Upon increasing the mass splitting, the cross section decreases at first due to suppression from an increased
propagator mass, until the longitudinal W contribution begins to dominate. The mono-jet cross section is monotonically decreasing with increase in propagator mass.}
\label{fig:crossx}
\end{figure}
 
In Fig. (\ref{fig:crossx}) we show the ratio of the cross sections for
the mono-$W$ and mono-jet processes at hadron level at the 14 TeV LHC,
as a function of $\lambda_4$.  (Although we have illustrated this
behavior for a particular choice of the $\chi$ and $\eta_u$ masses, we
obtain similar behavior for other parameter choices.) While the
mono-jet cross section monotonically decreases as $\lambda_4$ is
increased, caused by the heavier $\eta_d$ propagator, the mono-$W$
cross section first decreases and then increases again when radiation
of $W_L$ from the $\eta$ propagator takes over.  However, in order to
achieve a significant enhancement of the ratio of the mono-$W$ to
mono-jet cross sections, very large values of $\lambda_4$ are
required.  If we restrict this parameter to perturbative values,
$\lambda_4 < 4\pi$, a relative enhancement cannot be achieved.

This behavior differs greatly to that seen in $SU(2)$ violating EFTs,
where gauge non-invariant contributions from the analogue of
Fig. (\ref{fig:diags} a,b) lead to large $W_L$ contributions.  In our
renormalizable model, where all 3 diagrams of
Fig. (\ref{fig:diags} a,b,c) are properly included, the high energy
behavior of the cross section is tamed.

\subsubsection{$SU(2)$ Breaking and the $M_T$ Spectrum}

We now consider the $M_T$ distribution of the mono-$W$ events.  For
the EFT model of Eq.~(\ref{eq:xi}), the mono-$W$ $M_T$ distributions were
found to be sensitive to the parameter
$\xi$~\cite{Khachatryan:2014tva}. Compared to the $SU(2)$ conserving
choice $\xi=1$, the $SU(2)$ breaking choice of $\xi\neq 1$ resulted in a
harder $M_T$ distribution, with a higher peak and significantly more
high $M_T$ events.  This was useful in differentiating the signal from
background via appropriate cuts on the minimum value of $M_T$.

\begin{figure}[H]
  \centering
  \includegraphics[width=10cm]{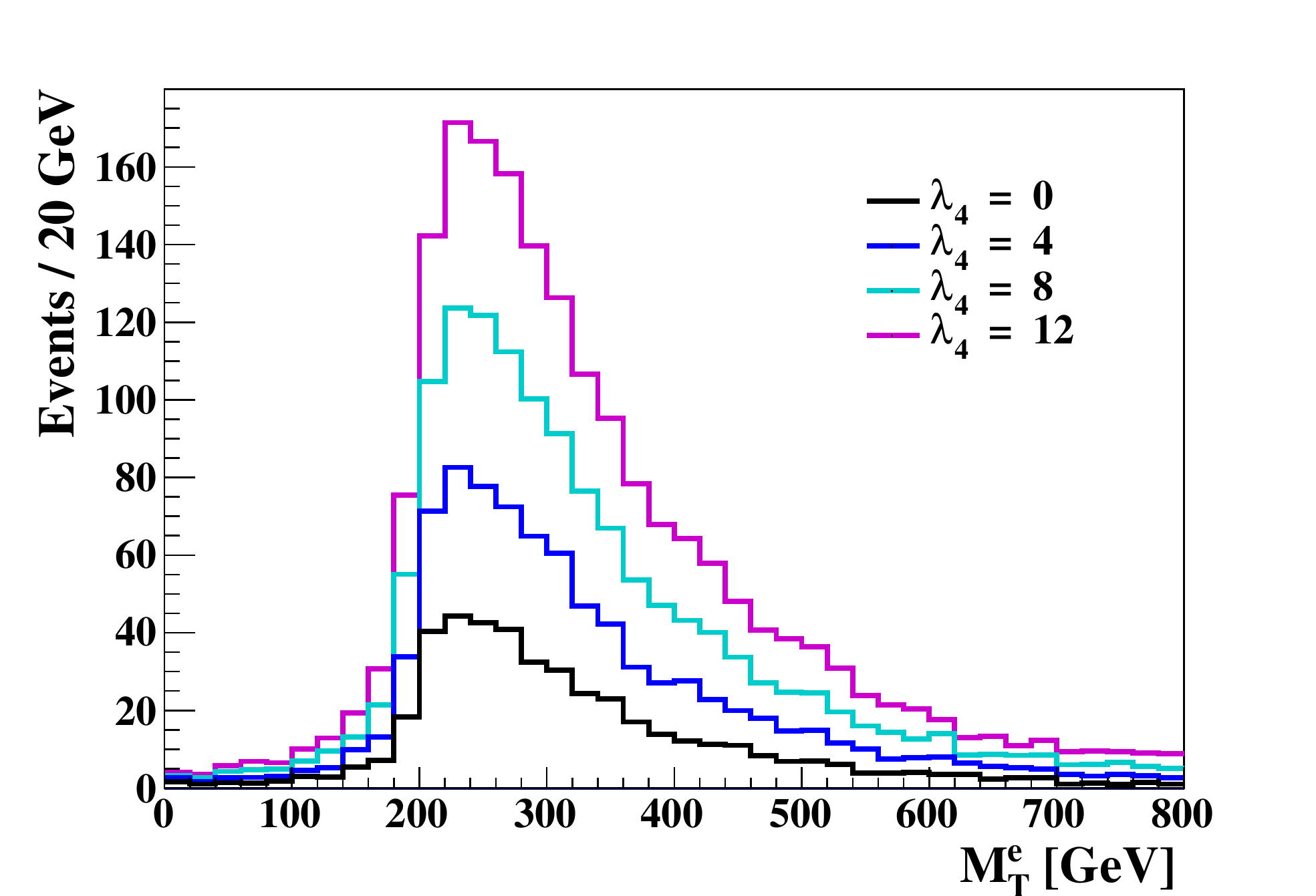}
  \caption{$M_T$ distribution for $m_{\eta_u}=220$ GeV, $g=1$, $m_\chi = 200$ GeV in the $t$-channel model with isospin violation, at 14 TeV and $\mathcal{L}_{int}=300$ $fb^{-1}$. Despite the increase in $\lambda_4$ and therefore the mass splitting, the peak of the $M_T$ distribution does not increase, leading
  to no strong advantage in the mono-lepton channel compared to other channels.}
\label{Fig:mtle}
\end{figure}

To explore this effect in our $t$-channel simplified model, we plot the
$M_T$ distribution for various choices of $\lambda_4$, shown in
Fig. (\ref{Fig:mtle}).  We see that increasing the mass splitting
parameter $\lambda_4$ produces no noticeable shift in the peak or
shape of the $M_T$ distribution.  Therefore, the shape of the $M_T$
distribution cannot be exploited to increase sensitivity.

\subsection{Isospin Violating Effects in $s$-channel Models}
\label{ZZ'}

We now consider $SU(2)$ violating interactions of DM with quarks in
the context of the $s$-channel $Z'$ model.  In the example model
presented in Section~\ref{sec:simplified}, the $Z'$ boson was taken to
couple with equal strength to the $u$ and $d$ type quarks.  This would
be expected in a scenario in which the SM quarks were charged under
the new $U(1)_{Z'}$.  However, if the SM quarks were not charged under
$U(1)_{Z'}$, and the $Z'$-quark couplings were to arise only via
mixing of the $Z'$ with the SM $Z$, then weak isospin violating
interactions would result -- see section A2 of
Ref.~\cite{Abercrombie:2015wmb}. In fact, these weak isospin
violating interactions would be the lowest order DM-quark interaction
terms present.

\begin{figure*}[ht]
\centering
\subfigure[]{\includegraphics[width=0.3\columnwidth]{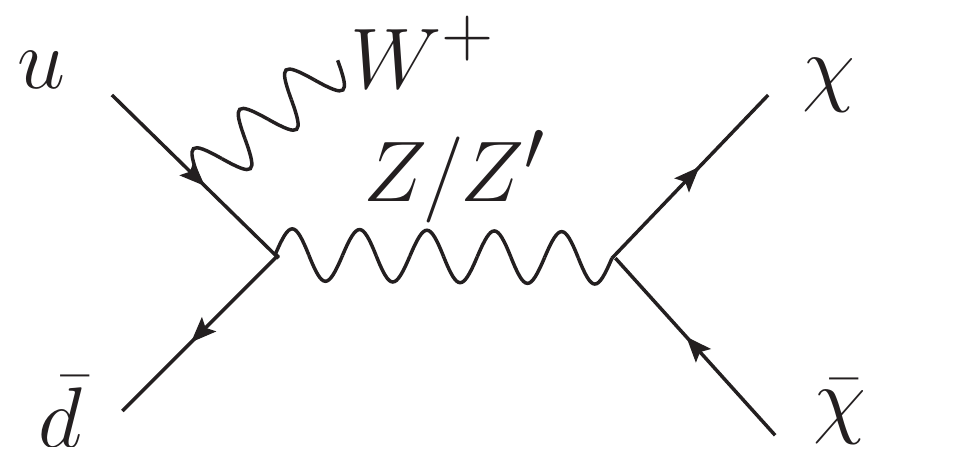}}
\subfigure[]{\includegraphics[width=0.3\columnwidth]{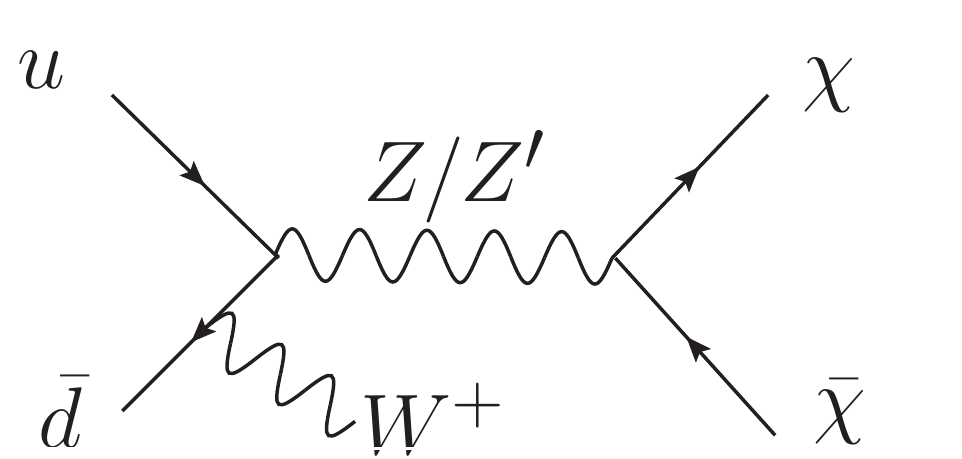}}
\subfigure[]{\includegraphics[width=0.3\columnwidth]{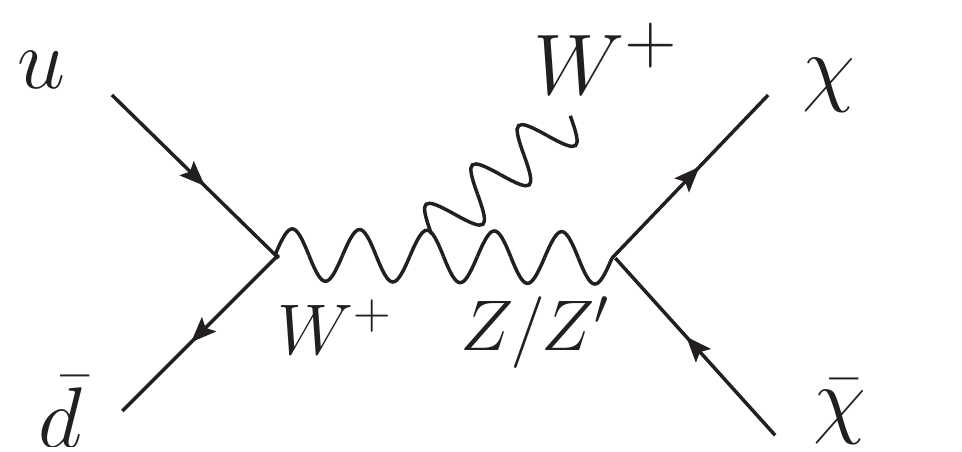}}
\caption{Contributions to the mono-$W$ process $ u(p_1)\overline{d}(p_2)  \rightarrow  \chi(k_1)\overline{\chi}(k_2)  W^+(q)$, in the $Z$-$Z'$ mixing model.}
\label{fig:diags2}
\end{figure*}

In the $Z$-$Z'$ mixing scenario the quark-$Z'$ couplings are
proportional to the quark-$Z$ couplings, which are of opposite sign
for $u$ and $d$ quarks due to their weak isospin assignments of
$T_3=\pm 1/2$.  In the EFT limit, where the $Z'$ is integrated out,
this would result in the operator of Eq. (\ref{eq:xi}) with a negative
value of $\xi$.  However, the strength of the DM-quark interactions
would be suppressed by the $Z$-$Z'$ mixing angle, which is of order
$v_{\rm EW}^2/M_{Z'}^2$ and thus the operator arises only at order
${1/\Lambda^4}$.  The relevant diagrams for the mono-$W$ process are
shown in Fig.(\ref{fig:diags2}).  Unlike the $Z'$ model of
Section~\ref{sec:simplified}, there is now a diagram in which the $W$
is radiated from the $Z/Z'$ mediator. This diagram occurs at the same
order in $1/\Lambda$ as the first two contributions\footnote{If we
  included only the first two diagrams, e.g., by assuming only the
  operator of Eq. (\ref{eq:xi}), we would encounter unphysical $W_L$
  effects whose origin could be traced to the lack of gauge
  invariance.}. While the third diagram will allow $W_L$ production, the
gauge invariance of the underlying theory prevents any bad high energy
behavior, limiting any $W_L$ driven cross section enhancement.
Moreover, given that the $Z$-$Z'$ mixing angle is constrained to be
small, isospin violating effects will be difficult to observe.

Finally, weak isopin violating effects would also occur in a model in
which a new $s$-channel scalar mediator mixes with the SM Higgs.  In
this case the effects are suppressed by the small SM quark Yukawa
couplings.  In addition, if the DM is lighter than the Higgs, the
Higgs invisible branching fraction would constrain the scalar-Higgs
mixing.

\section{Conclusion}

Observation of DM production at the LHC is now one of the foremost
goals of the particle physics community.  To analyze the sensitivity
of these searches, it is important to use a theoretically consistent
framework for describing the DM interactions.  The goal of this paper
was to explore mono-$W$ signals of dark matter production, in
simplified models in which invariance under the SM weak gauge
symmetries is enforced.  We therefore considered popular
simplified models with an $s$-channel $Z'$ mediator or a $t$-channel
colored scalar mediator, both with and without isospin violating
effects arising from electroweak symmetry breaking.

We first analyzed the simplified models in which the DM-quark couplings
preserve isospin.  Considering both the leptonic and hadronic decay
modes of the $W$, we found that the 8 TeV mono-$W$ sensitivity is not
competitive with the 8 TeV mono-jet results.  At 14 TeV the hadronic
(mono fat jet) decay channel is the most promising, although $3000$
$fb^{-1}$ of data is required to significantly probe parameter space.
While we anticipate that the experimental collaborations will be able
to better optimize their analyses than the estimates we present here,
we expect these general conclusions to hold.

Previous mono-$W$ analyses have focused primarily on EFT operators
that violate $SU(2)_L$, obtaining limits that are competitive with, or
stronger than, those arising from the mono-jet.  Therefore, we explored
the possibility of obtaining isospin-violating DM-quark couplings in
our gauge invariant simplified models, after electroweak symmetry
breaking.  This can be achieved in the $t$-channel model through the
mass splitting of the squark-like scalar $SU(2)$ doublet, or in the
$s$-channel model via $Z$-$Z'$ mixing.  For the both $t$-channel and $s$-channel
models we find that these isospin violating effects must be small, in
contrast to the non gauge invariant EFTs scenarios considered
previously in the literature.  As such, isospin violating DM-quark
couplings are unlikely to increase the sensitivity of mono-$W$
searches.

If DM is detected in future LHC data, it is likely that the mono-jet
process will be the discovery channel.  However, observation of a
mono-jet signal alone would not be sufficient to elucidate the
particular DM model.  Complementary information from other channels
such as the mono-$W$ would eventually play an essential role.
However, it will be challenging to observe these complementary signals
at the 14 TeV LHC unless the model parameters fall just beyond the 8
TeV mono-jet reach.  The observation of a mono-$W$ signal at the 14 TeV
LHC would therefore point toward very specific DM models.  While
mono-$W$ signals can, in principle, probe isospin violation of the
DM-quark couplings, encoding important information about the specific
DM model, it may take a future collider for such effects to be observed.

\section{Acknowledgements}
N.F.B., Y.C. and R.K.L. were supported by the Australian Research Council.  We acknowledge important
discussions with T.J.~Weiler and J.B.~Dent, and thank Y.~Bai for pointing out
the $Z$-$Z'$ mixing scenario of Section~\ref{ZZ'}. Feynman diagrams are made
using {\sc JaxoDraw} \cite{Binosi:2003yf}.

\bibliographystyle{JHEP}
\bibliography{fullmodel}


\end{document}